\documentclass[a4paper,11pt]{article}
\pdfoutput=1 

\usepackage{jinstpub} 
\usepackage{subfigure}
\usepackage{lineno}

\newcommand{\Deg}{\,^{\circ}}

\title{\boldmath A new method of energy reconstruction for large spherical liquid scintillator detectors}

\author[a]{W. Wu}
\author[b]{M. He}
\author[a,1]{X. Zhou\note{Corresponding author.}}
\author[a]{and H. Qiao}

\affiliation[a]{Hubei Nuclear Solid Physics Key Laboratory, Key Laboratory of Artificial Micro- and Nano-structures of Ministry of Education, and School of Physics and Technology, Wuhan University, \\ Wuhan 430072, China}
\affiliation[b]{Institute of High Energy Physics, Chinese Academy of Sciences,\\Beijing 100049, China}

\emailAdd{xiangzhou@whu.edu.cn}

\abstract{Liquid scintillators are commonly used to detect low energy neutrinos from the reactors, sun, and earth. It is a challenge to reconstruct deposited energies for a large liquid scintillator detector. For detectors with multiple optical mediums such as JUNO and SNO+, the prediction of the propagation of detected photons is extremely difficult due to mixed optical processes such as Rayleigh scattering, refraction and total reflection at their boundaries.
Calibration based reconstruction methods consume impractical time since a large number of calibration points are required in a giant detector. In this paper, we propose a new model-independent method to reconstruct deposited energies with minimum requirements on the calibration system. This method is validated with JUNO's offline software. Monte Carlo studies show that the energy non-uniformity can be controlled below 1\%, which is crucial for JUNO to achieve 3\% energy resolution.}


\begin{document}
\maketitle
\flushbottom


\section{Introduction} \label{sec.intro}

Liquid scintillator (LS) detectors have been widely used in modern neutrino experiments, such as KamLAND, Borexino, Double Chooz, Daya Bay and RENO \cite{Eguchi:2002dm,Bellini:2014uqa,Abe:2011fz,An:2012eh,Ahn:2012nd}.  KamLAND confirmed the large mixing angle solution of the solar neutrino oscillation \cite{Eguchi:2002dm}. Borexino observed the spectrum of proton-proton neutrinos in the sun \cite{Bellini:2014uqa}. Double Chooz, Daya Bay and RENO measured a nonzero value for the third mixing angle $\theta_{13}$ \cite{Abe:2011fz,An:2012eh,Ahn:2012nd}. Spherical LS detectors are popular for the next generation of neutrino experiments which will focus on the remaining problems such as the mass hierachy of neutrinos, solar neutrinos and neutrinoless double-beta decay. For instance, detectors of the on-going SNO+ experiment, the under-construction JUNO experiment and the planned Jinping neutrino experiment \cite{Andringa:2015tza,An:2015jdp,JinpingNeutrinoExperimentgroup:2016nol} are designed as spherical targets. In the SNO+ and JUNO detectors, acrylic is used to build containers for kilotonnes magnitude of liquid scintillator. Acrylic containers are surrounded by water buffers in order to shield external radioactivities \cite{Andringa:2015tza,An:2015jdp}. A similar detector configuration is taken into consideration as one option for Jinping \cite{JinpingNeutrinoExperimentgroup:2016nol}.

The energy reconstruction of individual events is a primary task in order to obtain neutrino information such as their energies and oscillation probabilities. In previous studies, event energies were reconstructed by a likelihood fit based on an analytical optical model \cite{Wen:2011ML} in which the temporal and spatial distributions of photoelectrons (PEs) are precisely calculated ignoring any scattering effect during photons' propagation. These distributions are smeared in a large detector at a scale of tens of meters when the photon scattering becomes significant. Moreover, they will be largely distorted by refractions and total reflections for the events at the edge of a detector because of different refractive indices between liquid scintillator and water. Therefore, it is challenging to precisely obtain the distributions of PEs by an optical model. Event energies can be also reconstructed with a vertex dependent correction related to the total number of collected photoelectrons based on calibration data, such as the method used in Daya Bay \cite{DYB:2017PRD}. For a spherical detector, in principle the collection of calibration data will be simplified to only along a radius because of the spherical symmetry. However, it is difficult to evenly distribute photomultiplier tubes (PMTs) on a spherical surface considering the procedure of the installation. Therefore the total charge response of events whose vertices have the same distance to the center of a spherical detector will not be uniform. As a result, the energy reconstruction method based on the vertex dependent corrections will lose the benefit of the spherical symmetry and needs hundreds of calibration points with precise positions. Therefore, it would be useful to find another way to derive the charge responses and then reconstruct individual event energies from the calibration data along a radius of a spherical detector.

In this paper, we have developed a new energy reconstruction method for spherical liquid scintillator neutrino detectors, which is described in section~\ref{sec.meth}. It has been studied by using Monte Carlo (MC) simulations and is validated to be useful for the detector configuration of JUNO in section~\ref{sec.perf}. Conclusions and discussions are discussed in section~\ref{sec.conc}.

\section{Methodology of the new reconstruction method}\label{sec.meth}

\begin{figure}[htbp]
    \begin{center}
        \includegraphics[width=0.5\textwidth]{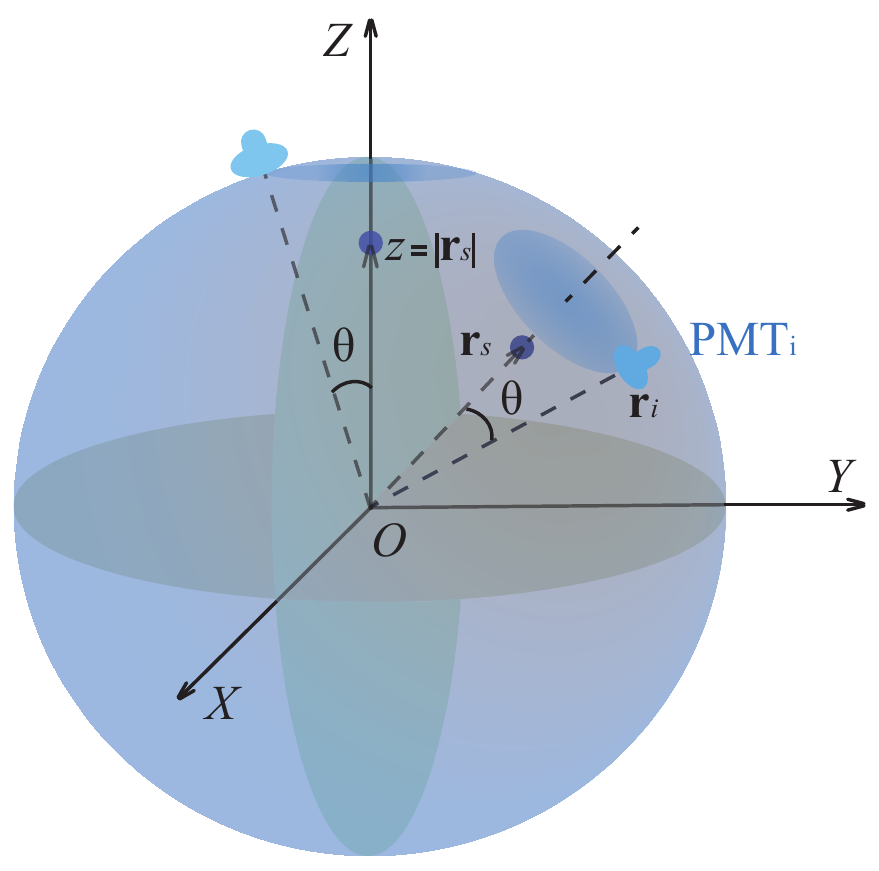}
        \caption{\label{fig:priciplediagram} Detector's coordinate system.}
    \end{center}
\end{figure}

The geometry of a spherical detector can be described in a global Cartesian coordinate system as shown in figure~\ref{fig:priciplediagram}.  The position of the $i$th PMT is $\mathbf{r}_{i} (x_{i}, y_{i}, z_{i})$.  Since PMTs are arranged at the same spherical surface, thus $|\mathbf{r}_i|=R$.  An event with mono-energy $E$ happens at an arbitrary position $\mathbf{r}_s(x, y, z)$. The distance between any position $\mathbf{r}$ and the detector center $O$ is $r=|\mathbf{r}|$. The angle between $\mathbf{r}_i$ and $\mathbf{r}_s$ is $\theta$. The registered number of photoelectrons (nPE) of the $i$th PMT is $n_{i}$, which depends on $\mathbf{r}_i$, $\mathbf{r}_s$ and $E$. Because of the spherical symmetry, $n_i$ can depend only on $r_s$, $\theta$ and $E$, i.e.
\begin{equation}
    n_i(\mathbf{r}_i,\mathbf{r}_s,E)=n_i(r_s,\theta,E).
\end{equation}
The number of detected photoelectrons, $n_i(r_s,\theta,E)$ has contributions from the absolute light yield, the attenuation of propagation mediums, the geometry effect and the detection efficiency of PMTs. In order to avoid the influence from the non-uniform installation of PMTs, we calculate the average nPE registered in the PMTs that have angle $\theta$ by
\begin{equation}
    \mu(r_s,\theta,E)=\overline{n_{i}(r_s,\theta,E)},
\end{equation}
where $\mu(r_s,\theta,E)$ is the mean nPE of those PMTs. If $Z$-axis is chosen as the symmetry axis to deploy a radioactive source at $\mathbf{r}_s(0,0,z)$ in order to obtain the nPE of the $i$th PMT $n_{i}(z,\theta,E)$, the response functions $\mu(z,\theta,E)$ is defined as
\begin{equation}
    \mu(z,\theta,E)\equiv\overline{n_{i}(z,\theta,E)}.
    \label{eq:responsefuncz}
\end{equation}
and thus it should satisfy
\begin{equation}
    \mu(z,\theta,E)|_{z=r_s}=\mu(r_s,\theta,E).
\end{equation}
It can also be applied to X-axis by analogy to deploy a radioactive source at
$\mathbf{r}_s(x,0,0)$ to obtain response functions defined as
\begin{equation}
    \mu(x,\theta,E)\equiv\overline{n_{i}(x,\theta,E)},
    \label{eq:responsefuncx}
\end{equation}
and so for the other axes. Any set of response functions obtained from one axis can be used as an input to the reconstruction algorithm.

The new reconstruction method is defined as the inverse operation of the above processes which is deriving the energy according to the distribution of collected photoelectrons, and $\mu(r_s,\theta,E)$ is the main tool to characterize the detector response and determine the event energy. Once the vertex of one event is known as $\mathbf{r}_s$, then the mean nPE collected by one PMT can be calculated. The probability function of the detected nPE for the $i$th PMT is following the Poisson distribution
\begin{equation}
    P(k_{i}|\mathbf{r}_s,E)=\frac{e^{-\mu_{i}}\cdot\mu_{i}^{k_{i}}}{k_{i}!},
    \label{eq:nPEPDF}
\end{equation}
where $\mu_{i}=\mu(r_s,\theta,E)=\mu(z,\theta,E)|_{z=r_s}$ is the expected number of photoelectrons for the $i$th PMT calculated from the response function.

A likelihood function $\mathcal{L}$ can be constructed with eq.~\ref{eq:likelihood} and if the detector is surrounded by $m$ PMTs.
\begin{equation}
    \mathcal{L}(k_{1},k_{2},\dots,k_{m}|\mathbf{r}_s,E)=\prod_{i=1}^{m}P(k_{i}|\mathbf{r}_s,E)=\prod_{i=1}^{m}\frac{e^{-\mu_{i}}\cdot\mu_{i}^{k_{i}}}{k_{i}!}.
    \label{eq:likelihood}
\end{equation}
It's an extensively used method to take the logarithm of $\mathcal{L}$ in order to replace continuous multiplications by continuous additions. Due to the monotonicity of logarithmic functions, it's more convenient to minimize $-\ln\mathcal{L}$ instead of maximizing $\ln\mathcal{L}$. Thus, the most probable value of $E$ is derived when $-\ln\mathcal{L}$ reach it's minimum.

\section{Performance studies of the new reconstruction method}\label{sec.perf}

\begin{figure}[htbp]
    \begin{center}
        \includegraphics[width=0.8\textwidth]{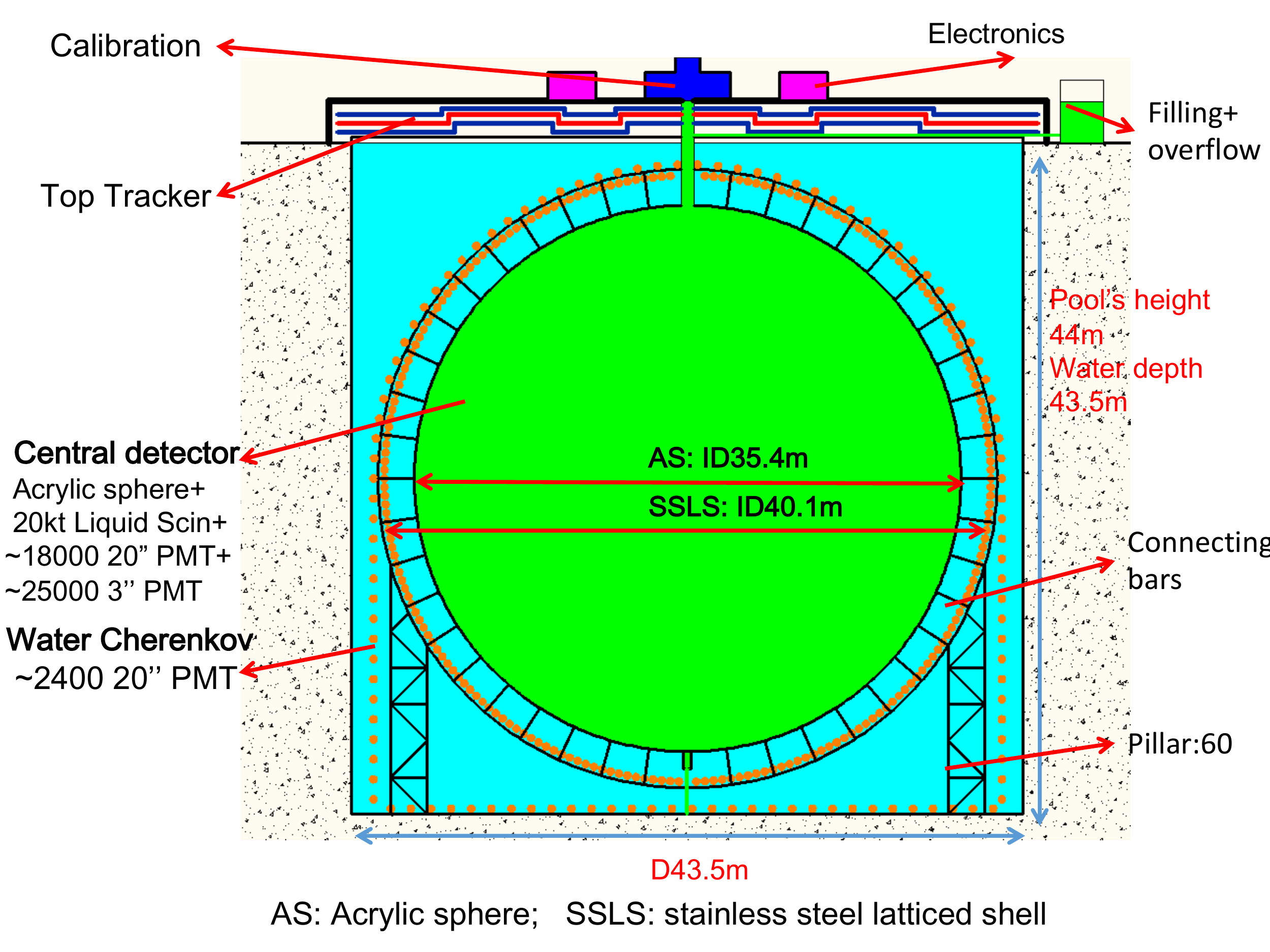}
        \caption{\label{fig:SchemaOfJUNO} Schematic diagram of JUNO.}
    \end{center}
\end{figure}

The central detector of JUNO is utilized in order to study the performance of the new reconstruction method. The schematic diagram of the JUNO detector is shown in figure~\ref{fig:SchemaOfJUNO}. There is 20 kiloton LS contained in an acrylic sphere with diameter of 35.4~m immersed in pure water, in which about 18,000 20-inch photomultiplier tubes (PMTs) and 25,000 3-inch PMTs are installed, providing larger than 75\% optical coverage \cite{An:2015jdp}. 

\begin{figure}[htbp]
    \begin{center}
        \subfigure[Mean nPE of one PMT as function of $\theta$ which is defined in section~\ref{sec.meth}. Red dots are calculated based on a rectilinear optical model. Blue circles are obtained directly from GEANT4 simulation in which absorption and reemission, Rayleigh scattering, refraction and the total reflection are considered.]
        {\label{fig:tradition.rec.a}\includegraphics[width=0.48\textwidth]{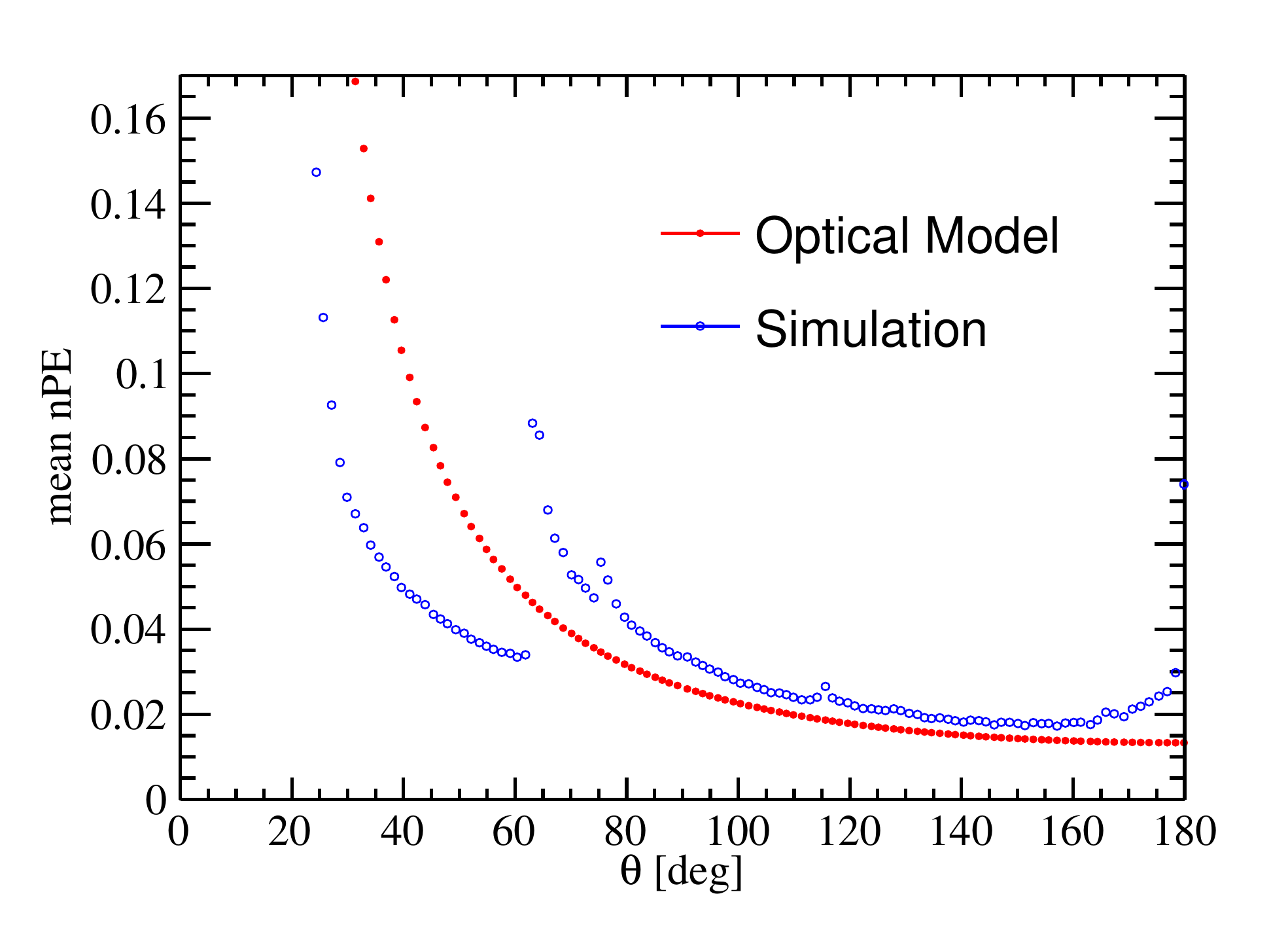}}
         \hspace{0.02\textwidth}
        \subfigure[Normalized total PE as a function
        of the event $z$-position at a fixed radius $R=16~\mathrm{m}$ ($\cos\theta_z=\frac{z}{R}$). The distribution of total PE is not uniform with respect to the positions which are located at a fixed radius. About 10\% non-uniformity is introduced due to the asymmetric arrangements of PMTs.]
        {\label{fig:tradition.rec.b}\includegraphics[width=0.48\textwidth]{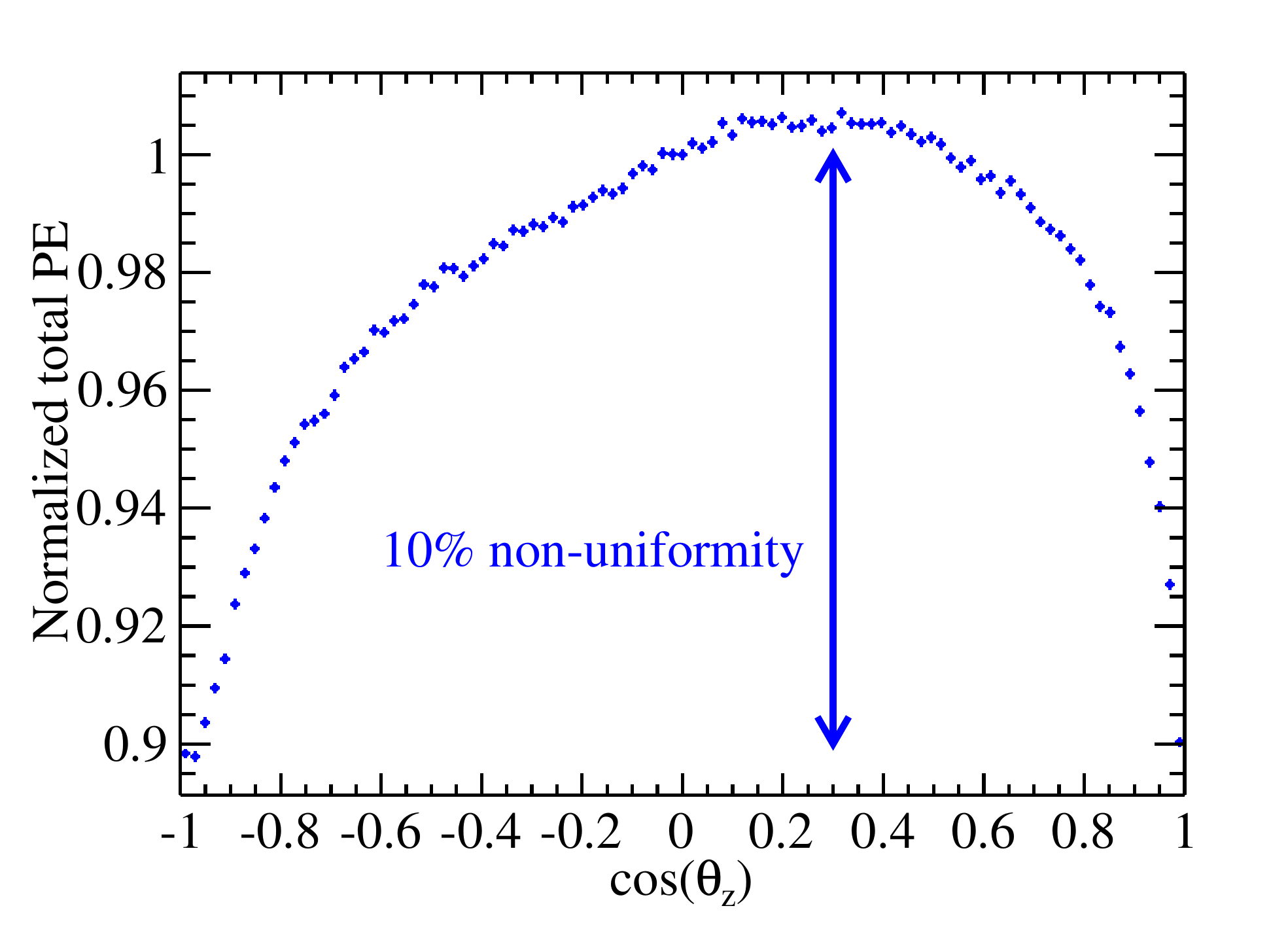}}
        \caption{Illustration effects impacting the propagation of optical photons and the response of the detector.}
        \label{fig:tradition.rec} 
    \end{center}
\end{figure}

JUNO is 20 times larger than any present liquid scintillator detector and the profile of expected nPE is deeply affected by absorption and reemission, Rayleigh scattering, refraction and the total reflection. These optical processes can be described with GEANT4 \cite{Geant4} simulation but only with precise knowledge of the optical parameters which need independent measurements. As a demonstration, a rectilinear optical model is introduced assuming all scintillation photons go straightly and the average nPE distribution is obtained and compared with the full simulation. In figure~\ref{fig:tradition.rec.a}, an example is shown for events located at (0, 0, 17) m. The rectilinear optical model predicts a smooth decrease of nPE with the increase of $\theta$ due to the acceptance angle of PMTs and the absorption of photons in the liquid scintillator, while the GEANT4 simulation gives discontinuous nPE distribution. The significant drop below $70\Deg$ comes from the total reflection at the edge of the liquid scintillator which shields part of PMTs from the light and the shielded region is called the ``dark zone''. For events located at (0, 0, 17) m,  the ``dark zone'' ranges from $29.9\Deg$ to $74.1\Deg$ for mono-wavelength light with 430 nm wavelength. This distribution is further smeared by the Rayleigh scattering and that is the reason why PMTs in the ``dark zone'' can still receive some photons. 
On the other hand, the complicated detector supporting structure and the nonuniform deployment of PMTs break the detector symmetry, as shown in figure~\ref{fig:tradition.rec.b}. Therefore, calibration of the detector response as a function of the event vertices is a heavy task. Generally speaking, the energy reconstruction with either the optical model or the event vertices dependent calibration is a challenge for JUNO.

In this study, only the 20-inch PMT system was used to test the new reconstruction algorithm that can be applied also to the 3-inch PMT system in principle. MC samples were generated in the JUNO offline software framework that was developed using the SNiPER software \cite{Lin:2017usg}, which contains the full detector geometry as shown in figure~\ref{fig:SchemaOfJUNO}. In particular, PMTs are arranged ring by ring from top to bottom. More than 500 PMTs are removed because of the connecting bars which support the acrylic sphere on the stainless steel latticed shell. There are 480 connecting bars implemented in the offline software currently. Each of them is constructed as a cylindrical hollow tube with 39 mm inner radius, 51 mm outer radius and 1200 mm height, which is made of steel in the simulation. They are connected to the outer surface of the central detector with copper nodes whose partial structures attached to the surface are cylindrical hollow tubs with 80 mm inner radius and 150 mm outer radius. Besides, a chimney with an inner diameter of 0.5~m extend up from the top of the sphere, serving as the filling port and the interface for the calibration system. Both the connecting bars and the chimney further break the symmetry of the detector's acceptance. 

Optical parameters of all propagation mediums are obtained from different measurements and implemented to SNiPER. For the scintillation process, the light yield of liquid scintillator is 11522 PE/MeV. The refractive indices are 1.496, 1.503 and 1.353 at 430 nm for LS, acrylic and water respectively.  The Rayleigh scattering length of LS is 27.02 m at 430 nm \cite{Zhou:2015gwa,Zhou:2015ps,Qian:2015NIM,Zhou:2015EPJC}. The Absorption length of LS is 79.61 m at 430 nm.  Both Rayleigh scattering length and absorption length are scaled at different wavelengths according to their spectra. The same quantum efficiency is applied to all PMTs, which is 29.38\% at 430 nm. Our new reconstruction algorithm is actually independent of optical parameters and the calibration template will be regenerated using the calibration source in the real detector where these parameters are unknown. 

\subsection{Performance studies in an ideal detector} 
The primary task of the new reconstruction method is to obtain response functions of the detector at different positions along one axis. In principle, an event vertex can be anywhere inside the detector. Considering a 1 m step-length, about 500 calibration points are required for filling a half cross section of the central detector. That calibration process needs about 3 days. In order to balance the precision and the detector live time, twenty-nine positions along one axis with different distances to the origin $r_s^{(j)}$ ($j$ = 1, 2, $\ldots$ , 29) are chosen to deploy the calibration source. The locations of deployment are from the origin ($r_s^{(1)}$ = 0 m) to the boundary ($r_s^{(29)}$ = 17.7 m) of the central detector. Intervals between two adjacent positions are 1 m from 0 to 16 m, 0.2 m from 16 to 17 m, and 0.1 m from 17 to 17.7 m respectively. In case the $r_s$ of an event doesn't equal to any one of $r_s^{(j)}$, the response function can be derived by taking a weighted average of response functions of two adjoining calibration points $r_s^{(k)}$ and $r_s^{(k+1)}$ which satisfy that $r_s^{(k)}$ < $r_s$ < $r_s^{(k+1)}$. The weight factor $\delta$ is calculated based on the distances between the vertex and corresponding calibration point. Thus the applied response function is calculated as
\begin{equation}
    \mu(r_s,\theta,E)=(1-\delta)\cdot\mu(r_s^{(k)},\theta,E)+\delta\cdot\mu(r_s^{(k+1)},\theta,E),
    \label{eq:AppliedRS}
\end{equation}
where $\delta=\frac{r_s-r_s^{(k)}}{r_s^{(k+1)}-r_s^{(k)}}$. Spline interpolation was investigated to calculate $\mu(r_s,\theta,E)$. Similar results were obtained with more computing time. Totally 50,000 events are generated at each position to suppress stochastic uncertainties to sub-percent level. For each event, 11,522 photons corresponding to the assumed light yield at the deposited energy of 1 MeV are produced isotropically, with wavelengths sampled according to the emission spectrum of LS.

\begin{figure}[htbp]
    \begin{center}
        \includegraphics[width=0.48\textwidth]{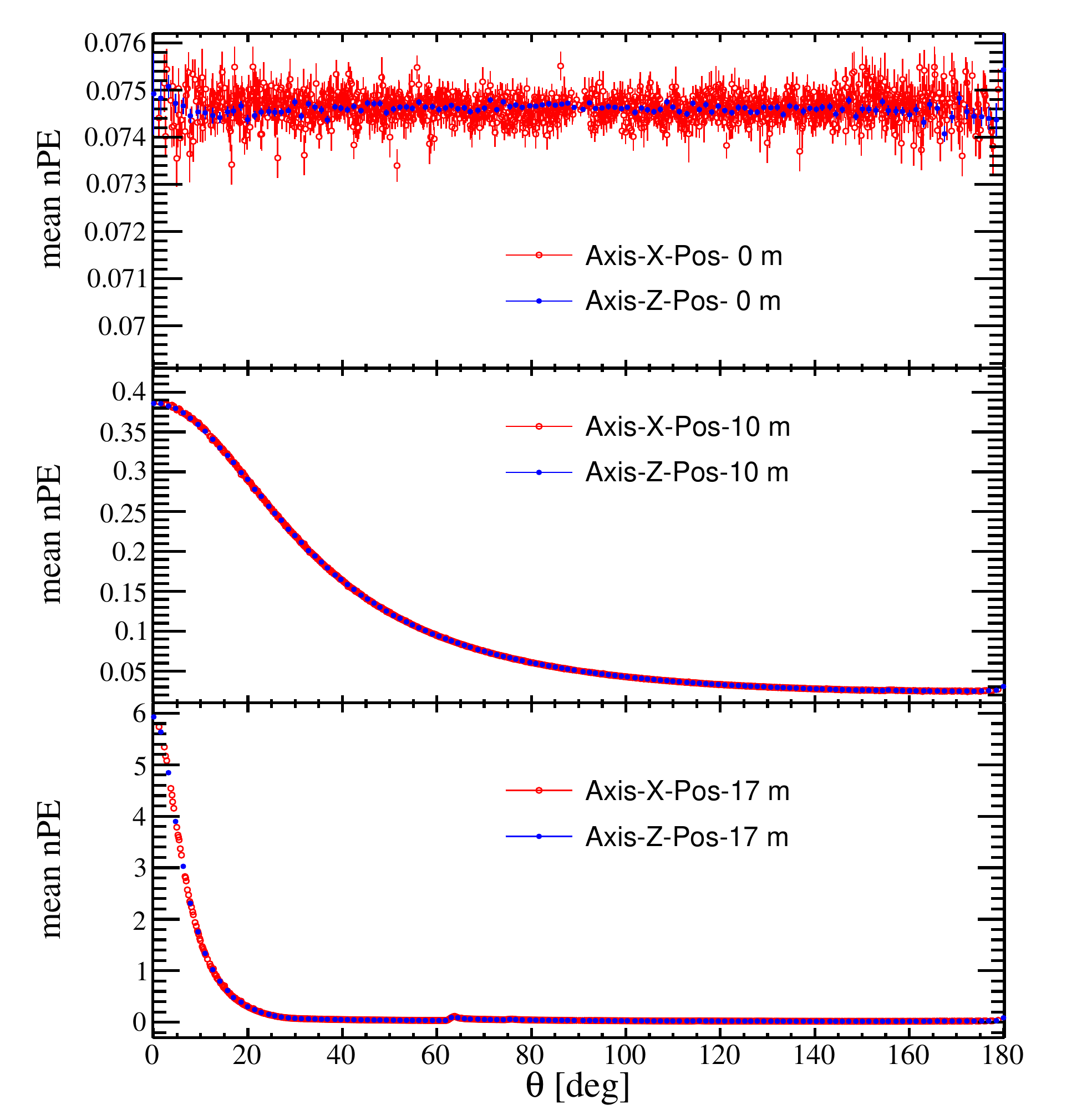}
        \caption{Two sets of response functions obtained along Z-axis (blue points) and X-axis (red point) at three illustrational positions for the ideal case that the connecting bars and chimney are not considered. Since PMTs are arranged by rings along the Z-axis, the blue points have less number of bins and smaller error bars for each bin. However, two response functions at each position are consistent within the margin of stochastic uncertainties.}
        \label{fig:RFWOWO}
    \end{center}
\end{figure}

In an ideal detector, the connecting bars and the chimney are not considered in the simulation. Response functions obtained along different axes are consistent with each other, thus one set of response functions along an arbitrary axis is enough for energy reconstruction. Two sets of response functions obtained along Z-axis and X-axis respectively are compared in figure~\ref{fig:RFWOWO} to validate the consistency of different axes. Three panels present response functions at three illustrational positions which are $r_s^{(0)} = 0~\mathrm{m}$, $r_s^{(11)} = 10~\mathrm{m}$, $r_s^{(22)} = 17~\mathrm{m}$ respectively. The two curves in each panel represent response functions obtained along Z-axis and X-axis separately which are $\mu(z,\theta,E)$ and $\mu(x,\theta,E)$ defined as eq.~(\ref{eq:responsefuncz}) and eq.~(\ref{eq:responsefuncx}) respectively. The two sets of response functions are well consistent with each other within the margin of stochastic uncertainties at different positions of the detector. Particularly, the two curves in the bottom panel are consistent while the ``dark zone'' effect appears. Therefore, response functions can be completely defined along a single axis.

\begin{figure}[htbp]
    \begin{center}
    \includegraphics[width=0.48\textwidth]{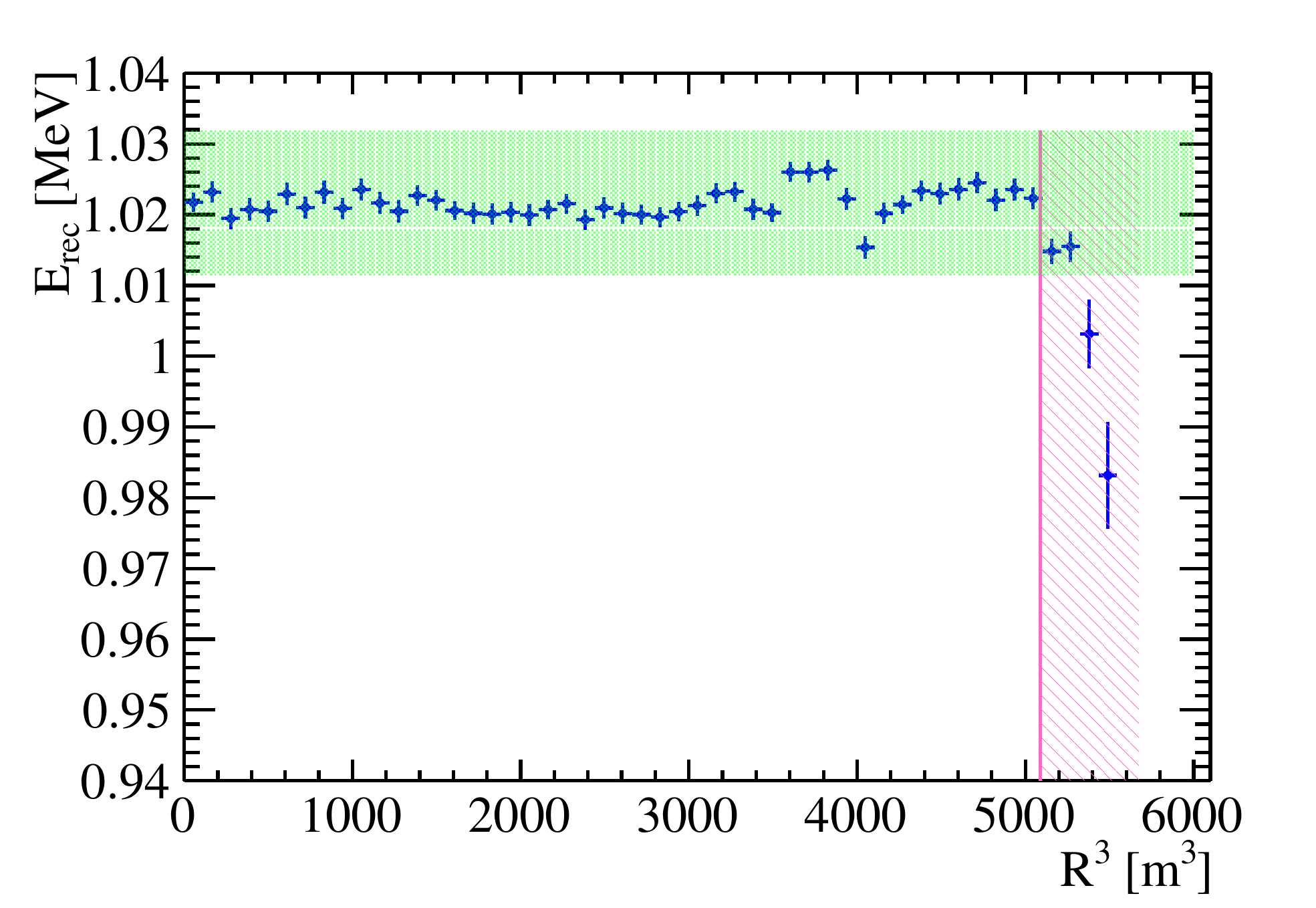}
    \caption{The energy non-uniformity is used to evaluate the performance of the reconstruction method. This figure shows the reconstruction result of static positron events in an ideal detector. Reconstructed energies in every volume bin inside the fiducial volume (left to the vertical pink exclusive zone) are falling into the 1\% band (horizontal green band).}
    \label{fig:UniWOWO}
    \end{center}
\end{figure}

Static positron events are uniformly generated in the LS region and reconstructed using response functions obtained along X-axis. In order to decouple energy reconstruction from vertex reconstruction, true vertices were extracted from simulation and used as input parameters for energy reconstruction. The released energy of each event is 1.022 MeV which is the sum of the stationary energy of a positron and that of a corresponding annihilated electron. A fiducial volume of 18.35 kt corresponding to a radial cut of 17.2~m was chosen since it is particularly effective in reducing the accidental background, mostly arising from $^{238}$U/$^{232}$Th/$^{40}$K contamination from the acrylic vessel, PMT's glass, steel supports, and copper fasteners \cite{An:2015jdp} . Therefore, events outside the fiducial volume are ignored for this analysis based on true positions. The energy non-uniformity describes the fluctuation of reconstructed energies with respect to the positions of their vertices and it is used to evaluate the performance of the reconstruction method. The non-uniformity is an important factor that affects the energy resolution as a non-statistical contribution \cite{Djurcic:2015vqa}. The result is presented in figure~\ref{fig:UniWOWO}. The vertical pink exclusive zone is outside the fiducial volume defined above. And the horizontal green band is the $\pm1\%$ region around the average reconstructed energy inside the fiducial volume. 

\subsection{Performance studies in a fully assembled detector}

\begin{figure}[htbp]
    \begin{center}
        \includegraphics[width=0.48\textwidth]{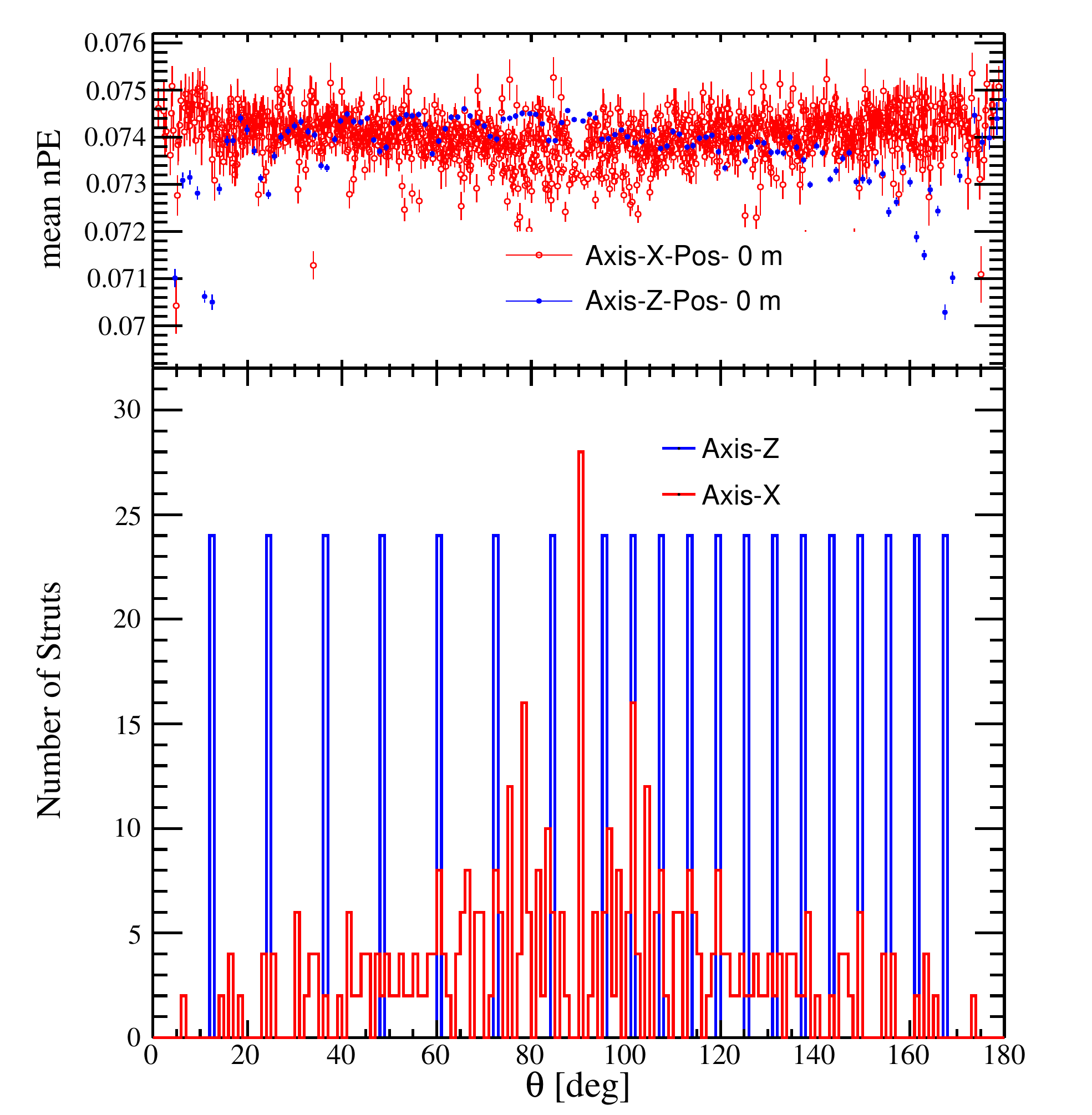}
        \caption{The top panel shows two sets of response functions obtained along Z-axis and X-axis at the center of the detector for the case in which connecting bars are considered. The bottom panel shows the arrangements of connecting bars in the simulation. The angle $\theta$ is defined in figure \ref{fig:priciplediagram}.}
        \label{fig:RFWW}
    \end{center}
\end{figure}

A fully assembled detector is not perfectly symmetrical due to the existence of connecting bars and the chimney that are necessary for mechanical requirements and calibrations as in JUNO. The top panel of figure~\ref{fig:RFWW} presents two sets of response functions obtained from X-axis and Z-axis respectively for the case in which connecting bars and the chimney are assembled. The bottom panel presents the arrangement of connecting bars with respect to different axes. The blue curve in the top panel has 20 distinct dips at the corresponding angles at which connecting bars are installed as the blue curve in the bottom panel shows. The red curves have similar correspondence with less significance. The distortion of both response functions in figures \ref{fig:RFWOWO} and \ref{fig:RFWW} shows that connecting bars have shadow effects on the nearby PMTs and therefore the reconstruction results will be consequently worse. In addition, the mean nPE obtained from Z-axis in figure \ref{fig:RFWW} drops at the very small angle due to the existence of the chimney. If response functions obtained from Z-axis were applied, the reconstruction capability would be damaged. Nevertheless, this drop can be easily corrected as long as an axis other than Z-axis is selected since the chimney has little influence on the average nPE of PMTs at the same angle, e.g. the X-axis as shown in figure \ref{fig:RFWW}. Therefore, the reconstruction is robust in an imperfectly symmetric detector.

\begin{figure}[htbp]
    \begin{center}
        \subfigure[Positron events with 0 MeV/c momenta.]
        {\label{fig:UniWW}\includegraphics[width=0.48\textwidth]{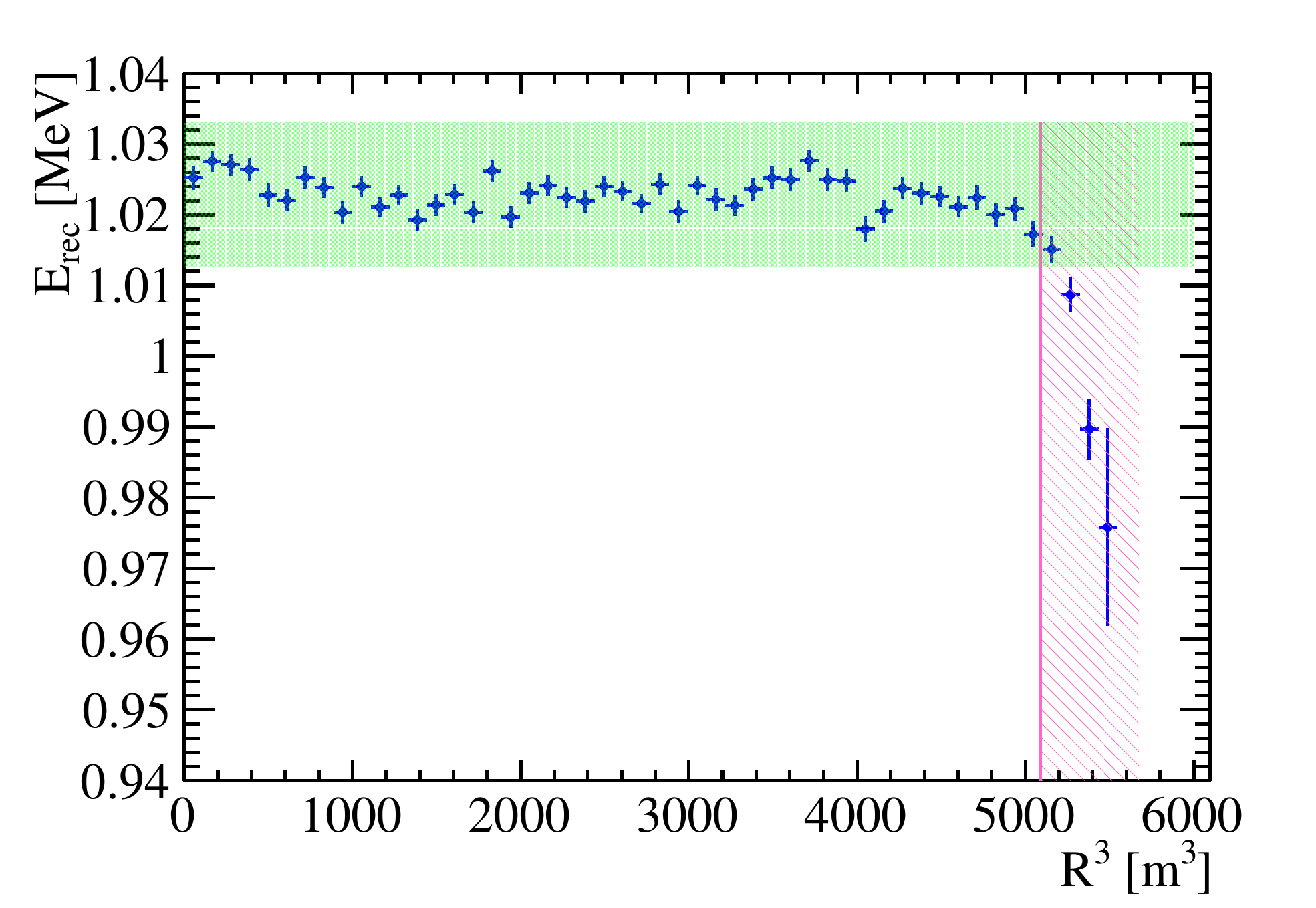}}
         \hspace{0.02\textwidth}
        \subfigure[Positron events with 7 MeV/c momenta.]
        {\label{fig:UniWW7MeV}\includegraphics[width=0.48\textwidth]{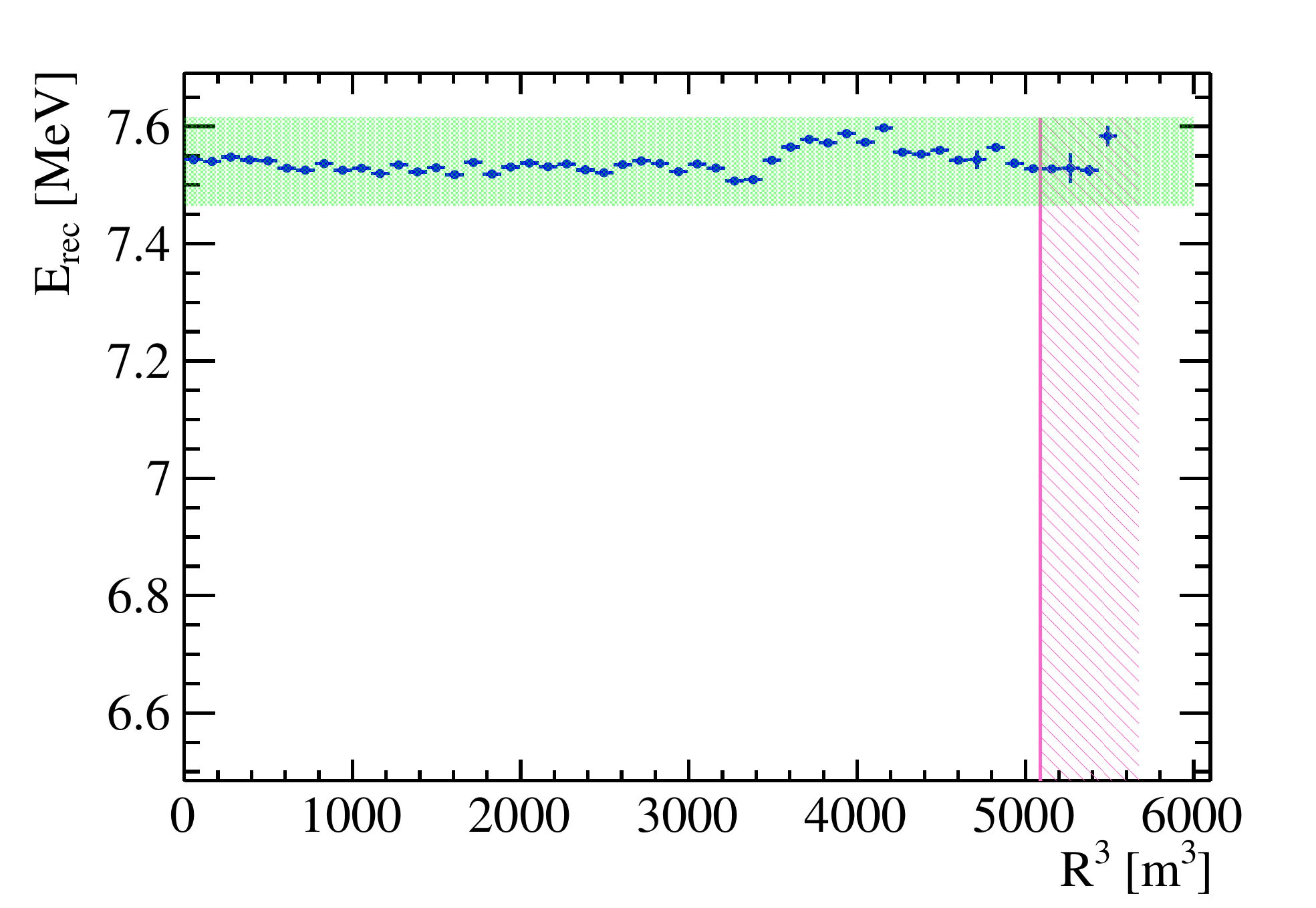}}
        \caption{Reconstruction results in the fully assembled detector. Reconstruction energies in every volume bin inside the fiducial volume (left to the vertical pink exclusive zone) are falling into the 1\% band (horizontal green band).}
        \label{fig:UniWWs} 
    \end{center}
\end{figure}

Positron events with momenta from 0 to 7 MeV/c were uniformly generated in the LS region and reconstructed using response functions obtained along X-axis. Besides true vertices, Gaussian distributed vertices around the true value with standard deviations from 0 to 20 cm were used as input parameters for the energy reconstruction. The non-uniformity is controlled within 1\% in any case, which shows our algorithm is not sensitive to the event vertex smearing at a reasonable range and can be applied to the major part of the reactor antineutrinos energy. The result of static positron events with true vertices is presented in figure~\ref{fig:UniWW}, which shows the non-uniformity of reconstructed energies in a fully assembled detector is better than $1\%$ within the fiducial volume. Similarly, the results of positron events with 7 MeV/c momenta is presented in figure~\ref{fig:UniWW7MeV} which gives the same conclusion.

\section{Conclusions and discussions}\label{sec.conc}
In this paper, a new energy reconstruction method is developed for large spherical liquid scintillator detectors, especially when there are multiple optical mediums with different reflective indices. Initially, the method is validated with an ideal central detector of JUNO in which the connecting bars and chimney are not considered. MC studies show that it can achieve sub-percent non-uniformity with minimum requirements on the calibration system. 50,000 events at each calibration position were produced to obtain response functions. Based on 100 Hz emission frequency of $^{60}$Co which is one of the radioactive sources used in Daya Bay \cite{Liu:2013ava}, 8.3 minutes are necessary to finish one calibration point and 4 hours are needed to finish the whole calibration procedure neglecting the moving time of the source temporarily. The cost time is comparable to that in Daya Bay which is 3 hours, so the calibration is cost-effective considering JUNO's 200 times larger volume. 

In addition, the method is applied to the fully assembled central detector of JUNO. The spherical symmetry is broken in the realistic construction of the detector due to the non-uniform installation of connecting bars and the chimney. The connecting bars affect the performance of the energy reconstruction not only by its symmetry breaking effect, but also by decreasing the total nPE due to the shadow effects. Besides, the chimney on the top of the central detector makes the response functions obtained along Z-axis incomplete. Nevertheless, these effects are largely reduced by averaging the PMTs at the same angle when the calibration axis moves from Z-axis to any of others such as X-axis. The method is validated for events with different energies and vertex resolutions. The reconstruction results show that the new algorithm is robust in order to achieve sub-percent non-uniformity. 

The new reconstruction method is generally developed for spherical detectors which are widely applied in modern neutrino experiments. In order to take into consideration every propagation processes of photons in different mediums, response functions obtained directly from calibration data along one axis are used to calculate the expected nPE of each PMT. The essential advantages are the spherical symmetry of the detectors and the averaging operation on the response functions, which eliminates the non-uniform geometrical effects such as the asymmetrical positioning of PMTs and connecting bars. Hence, enormous calibration data are avoided other than the reconstruction with a vertex dependent correction. In addition to JUNO, the method is also useful for other experiments with spherical detectors such as SNO+ and Jinping.


\acknowledgments
This work was supported by National Natural Science Foundation of China (Grant No. 11575226), the Major Program of the National Natural Science Foundation of China (Grant No. 11390381), the Strategic Priority Research Program of the Chinese Academy of Sciences (Grant No. XDA10010900, XDA10011200), and the Double First Class University Plan of Wuhan University. We sincerely thank Dr. Nikolaos Vassilopoulos for proofreading this article.



\end{document}